# Topological Derivation of Shape Exponents for Stretched Exponential Relaxation

J. R. Macdonald[1] and J. C. Phillips[2]

1. Dept. of Physics and Astronomy, University of North Carolina, Chapel Hill, NC, 27599-3255
2. Dept. of Physics and Astronomy, Rutgers University, Piscataway, N. J., 08854-8019

In homogeneous glasses, values of the important dimensionless stretched-exponential shape parameter $\beta$ are shown to be determined by magic (not adjusted) simple fractions derived from fractal configuration spaces of effective dimension $d^*$ by applying different topological axioms (rules) in the presence (absence) of a forcing electric field. The rules are based on a new central principle for defining glassy states: equal *a priori* distributions of fractal residual configurational entropy. Our approach and its $\beta$ estimates are fully supported by the results of relaxation measurements involving many different glassy materials and probe methods. The present unique topological predictions for $\beta$ typically agree with observed values to ~ 1% and indicate that for field-forced conditions it should be constant for appreciable ranges of such exogenous variables as temperature and ionic concentration, as indeed observed using appropriate data analysis. The present approach can also be inverted and used to test sample homogeneity and quality.

# 1.    Background

Among Nature's nearly infinitely many materials one finds only a few good glass formers.  The rarity of such materials is not widely appreciated, because the attention of science has focused on the few that do exist (like window glass) because of their immense technological value.  Almost all materials are "normal", in the sense that when they are cooled slowly their viscosity decreases linearly with temperature, until the freezing point is reached, where they crystallize.  Good glass formers, by contrast, encounter configurational barriers to crystallization on laboratory time scales.  As a result, they can be cooled far below their equilibrium freezing points into metastable states characterized by an exponentially increasing viscosity $\eta$ described [1] by ubiquitous functions of the Vogel-Fulcher form $\eta \sim \exp([T^*/(T - T_g)]^{\alpha})$ with $\alpha \sim 1$.

At present there is no consensus regarding the nature of the configurational barriers to crystallization that characterize good glass-forming liquids.  There are many families of good glass formers (network glasses, such as window glass; molecular glasses, such as glycerol; polymeric glasses, such as synthetic rubber, fused salts, …, and even electronic glasses) [2], and it seems likely that each family utilizes different kinds of configurational barriers to avoid crystallization.  There is one practical observation that can be made: whatever these barriers are: it seems likely that they can be studied more easily, more accurately, and more completely in the glass than in the supercooled liquid.  However, as implied by the Vogel-Fulcher relation, the viscosity itself is so large in glasses as to be non-measurable.

Fortunately, experiment has shown that in glasses that are microscopically homogeneous there is a second ubiquitous relation that describes relaxation of any probe pulse of intensity I as a function of time: $I(t) \sim \exp[(-t/\tau)^{\beta}]$, where conventional exponential relaxation with a relaxation time $\tau$ is obtained when the dimensionless fraction $\beta = 1$.  In fact, this function (called stretched exponential relaxation (SER)), also describes relaxation in metastable glass-forming liquids with $\beta < 1$. As T decreases towards $T_g$, $\beta$ decreases, sometimes reaching a plateau near or somewhat above $T_g$, depending on the probe. Glassy plateau values of $\beta$ for field-free relaxation are discussed in great detail in a lengthy 1996 review [2]; it is also an important subject of this paper for a wider range of experimental probe methods and materials, including both field-free and field-forced relaxation behavior.

One can "explain" SER by noting that it corresponds just to a particular distribution of relaxation times, and one would expect the relaxation times of glasses and supercooled liquids to be broadly distributed, because of the exponential complexity of the Vogel-Fulcher relation. This explanation, however, begs the question of why in microscopically homogeneous glasses and supercooled liquids the distribution of relaxation times *always* has the stretched exponential form or one directly derived from it, and it does not indicate a path towards understanding and explaining observed values of the *dimensionless* shape parameter $\beta$.

Broadly speaking, at present one can distinguish two different ways of approaching descriptions of the glassy state. One can start with the picture of a normal liquid with independent gas-like particles, and add correlations through polynomial interactions from presumed collective modes (as in mode-coupling theory). These hydrodynamic mode-coupling models are useful, for example, in weather forecasting, and they can be used to fit some properties of the supercooled liquid, especially after three-mode interactions are added [3] that include adjustable parameters to fit $T_g$ (poorly predicted by the two-mode model). Although this polynomial approach is useful for curve-fitting of some kinds of experimental data (especially scattering data taken by the very powerful spin-polarized neutron method [2,3]), it still leaves unanswered the central question discussed here, namely the meaning of $\beta$.

In this paper, and in the lengthy review [2], one takes a different approach. We argue that the appearance of the singular (or stretched) exponential functions in supercooled liquids (or glasses) shows that no polynomial, perturbative, or hydrodynamic approach based on ideas appropriate to normal gases and liquids is appropriate for glasses. Instead, one utilizes axiomatic topological models that are qualitatively different from geometrical models of normal liquids. These topological models plausibly assume that glasses, which differ from normal liquids because they are "frozen" into configurations with exponentially large (effectively infinite) viscosities, correspondingly occupy exponentially small fractions of free particle configuration space. Thus all geometrical real-space models, especially those that depend on parameterized polynomial assumptions, are exposed to the risk that the configurations they describe are not, in fact, *consistently* glassy (exponentially entangled). As an aside, one should note that this kind of problem has been familiar to mathematicians, especially set theorists and topologists, for more than 100 years; they refer to such problems as "Non-Polynomial (NP) complete", or exponentially complex. The prototypical exponentially complex problem is that of the traveling salesman who should visit randomly placed cities along the shortest path. Mathematicians frequently emphasize that such problems cannot be solved by brute-force polynomial methods, not even with the largest computers.

One of the most obvious weaknesses of many theoretical papers is that they are connected to only one (more often, no) experiment. Moreover, older SER experiments (say, before 1980) were often made with samples that were not microscopically homogeneous; in addition, relaxation measurements require large data bases, and before the widespread use of digital data acquisition and processing, these older data bases, many incorporated in the broad collection of [4], contained many errors. More recent digital data have been of much higher quality, and these were identified and emphasized in the lengthy review [2]. Moreover, that review predicted certain "magic" values of $\beta$, and these predictions have been confirmed (with accuracies of order a few or even one per cent), often without the authors being aware of the predictions, in *almost all* subsequent experiments. The *predictive success* of the present approach in analyzing experimental data could be taken as paradigmatic for quality control of analysis of a large, broad range of experimental data. Several new examples are discussed below.

Although older aspects of the topological approach have been reviewed in depth [2], it is summarized here for the reader's convenience. All glassy relaxation is supposed to take place in an exponentially restricted configuration space that is drastically different from conventional free particle configuration space. There are many ways that could be used to describe this abstract space, but happily it has turned out that the simplest way is by far the most successful, as verified by comparison with many experiments [2], and subsequent amazingly successful predictions (described here). The simplest way is topological: one imagines that relaxation paths are obtained by diffusion to relaxation sinks or defects [1,2] in a space of possibly restricted dimensionality $d^* \leq d = 3$. There are a very few simple axiomatic rules for determining $d^*$, which turns out to be of the form $fd$, with $f = m/n$, where $m$ and $n$ are *integers* (not irrational numbers!) and $m < n < 10$. In other words, the configuration space is fractal, but it is a very simple fractal space. The integers $m$ and $n$ arise in obvious physical ways, and they are not adjustable; there are only a few possibilities, and once determined for a given material, process, and probe, they are fixed and unique for the given situation.

There are two kinds of data on glass SER: temporal data, where the relaxation time $\tau$ and $\beta$ can be obtained simply by fitting the residual ($t > 100\tau$ [2]) field-free relaxation regime, and dispersive frequency data, usually obtained by measuring steady-state dielectric (field-forced) responses over limited frequency ranges [2,4,5]. Because glassy states occupy only exponentially small fractions of phase space, one should not assume (as is usually done) that the values of the dimensionless stretching fraction β are the same when measured in steady-state dielectric, ($\beta_f$) and in field-free ($\beta_0$) relaxation experiments. Standard procedures exist for obtaining $\tau$ and β in the former field-forced case by fitting to appropriate numerical frequency-response models, but these must be handled with exceptional care to obtain reliable results, especially for β [5,6]. The kind of SER that is observed also depends sensitively on the excitation and measuring probes used, and it is much more sensitive to the nature and scale of sample inhomogeneities (static or dynamic). Usually there is independent evidence of the existence of static inhomogeneities. Such evidence is increasingly plentiful and may be well known to experts in specific materials [2], but it is often overlooked in global geometrical models that claim to explain the microscopic origin of SER.

To establish the *intrinsic* behavior of SER in homogeneous glasses we have critically reviewed and augmented the very large (but old) data base [4] of temporal SER structural relaxation experiments, ones that cover a wide range of undifferentiated homogeneous and statically inhomogeneous materials, probes, and detectors [2].  We have also critically deconvolved doped-dielectric (conductive-system) dispersion data, being careful to separate the dispersive (ionic or dopant) part from the non-dispersive host electronic and dipolar parts, and adequately accounting for the effects of charging at electrode interfaces [5,6].

Our results may be summarized as follows: there exists a surprisingly large number of microscopically homogeneous glasses, including the most important ones technologically, whose properties, including the dimensionless stretching factor $\beta$, have been established with great accuracy ($\sim 1\,\%$ in $\beta$ in many cases), through careful analysis. The data exhibit simple "magic" $\beta$ fractions that can be explained *entirely topologically*, that is, in terms of only the connectivity and dimensionality of relaxation paths in configuration space.  Geometrical factors, such as filling factors [7], or ranges of interactions [8], become important only in inhomogeneous glasses.

Moreover, the topological dependence of the "magic" $\beta$ fractions on the effective dimensionality $d*$ is different in the presence of a forcing electric field involving mobile charge carriers (conductive-system relaxation, $\beta_f$ [6]) than it is for field-free structural relaxation ($\beta_0$); the differences, though small, are systematic, independent of microscopic geometry, and fully explicable.  **To the best of our knowledge, there exist at present no alternative parameter-free models that have successfully identified and predicted *intrinsic* values of the dimensionless stretching fraction $\beta$ in either regime.  The validity of the present topological approach is confirmed by its many successful predictions discussed below.**

## 2. Field-free Relaxation Examples

While much has been said rather generally about glass inhomogeneities, one can appreciate their significance most easily through examples. Silica, $(Na_2O)(SiO_2)_4$, and $(Na_2O)_{16} (CaO)_{10}(SiO_2)_{74}$ (window glass) are all homogeneous network glasses, but $GeO_2$ is not (in addition to the host four-fold coordinated Ge, it contains inclusions of five- or six-fold coordinated Ge in the form of deformed nanocrystallites with the much denser rutile structure [2]). Thus $\beta_0$ (silica, $(Na_2O)(SiO_2)_4$) = 0.60, (the theoretical magic fraction 3/5), while $\beta_0 (GeO_2)$ = 0.9; the former value is intrinsic, while the latter is not [2]. Similarly, polybutadiene (synthetic rubber) is a prototypical polymer; properly prepared samples studied as a function of vinyl (cis-trans) sidegroup fraction gave a nearly constant $\beta_0$ = 0.41-0.43, in agreement with the predicted topological magic fraction 3/7 [2], whereas varying the vinyl fraction in samples stabilized by peroxide addition gave the much wider range 0.50-0.26; the former nearly constant values are nearly intrinsic and independent of geometrical factors, while the latter widely varying ones are not (see Fig. 20 of the review [2]). The "worst case" polymer, polyvinyl chloride (PVC), is partially crystallized [2].

One of the most accurate ways of measuring $\beta_0$ is through stress relaxation, which can take place at temperatures low compared to the glass transition temperature, and over very long times (see Fig. 15 of the review [2]). This method has been used for several chalcogenide network alloy glasses, including pure Se itself (chain structure, few rings, unlike S), with results in excellent agreement with theoretical predictions [2]. The stress results for Se, measured on a time scale of $10^3$ s, concur exactly with those obtained by neutron spin-echo studies with a time scale of $10^{-9}$ s, and they also agree with the predicted theoretical value of $\beta_0 = 3/7 \simeq 0.43$. Exact concurrence may be fortuitous, but as theory and the two experiments all appeared concurrently and independently, it is unreasonable to dismiss such concurrence as the result of data selection.

More recently, the relaxation of polystyrene (PS) has been studied with stress relaxation [9]. PS consists of the usual hydrocarbon main chain with a phenyl side group. Pure PS is atactic (shapeless, presumably because of buckling stresses caused by the bulky side group) [2]. Commercially PS is stabilized, and presumably rendered microscopically homogeneous, by the addition of a suitable volume fraction of acrylonitrile. How can one test this picture of the commercial product, an exponentially unlikely "straw" found as the result of a searching through a very large "haystack" by trial and error? It turns out that analysis of the birefringence of commercial poly(styrene-*co*-acrylonitrile) films [9] led to the identification of two relaxation processes, that of the main chain, and that of the phenyl side group. The main chain shows the same value of $\beta_0 = 3/7 \simeq 0.43$ as Se chains did in both experiments described above, while the phenyl side group gave $\beta_0 = 0.32$. The interpretation of the side group result will be given below after discussion of the microscopic axiomatic model, but it strains one's credulity to believe that all these concurrences are merely the result of data selection, as is sometimes suggested without statistical or factual support.

Among non-polar molecular glass formers, orthoterphenyl (OTP) is the most studied, because it is available commercially in high purity. Relaxation in OTP has been studied by very accurate multidimensional (spin echo) NMR methods [10], yielding two values of $\beta_0$ for long and short preparation times, 0.59 and 0.42, compared to the topologically *predicted* fractions [2] 3/5 and 3/7, respectively. Brillouin scattering experiments also involve both long-and short-range forces, and these also give $\beta_0 = 0.43$ for OTP [11]. Relaxation of propylene glycol (PG) molecules and polymers (PPG), where relaxation is driven by OH groups, vicinal in PG, but separated in PPG by 70 non-dipolar monomers [12], shows $\beta_0 = 0.43$ (PPG) and 0.6 (PG), agreeing *exactly* with topological *predicted fractions* [2]. Similarly good agreement (~1%) with *prediction* is obtained for orientational glasses, both dipolar [2] and quadrupolar [13]. The doped (conductive-system) dielectric relaxation examples listed in Table I show virtually identical values of $\beta_f$ [5]: this similar behavior is also explained below topologically, *without the use of adjustable geometrical parameters.*

### 3. Temporal Trap Models

By now we expect that most readers will have been impressed enough with the examples above of spectacular agreement between theory and experiment to want to know briefly how the theory (described in detail in the lengthy review [2]) actually works. Many derivations of structural SER begin with an axiomatic trap model similar to that introduced in the seminal context of dispersive transport [14]. In the field-free case, excitations are supposed to diffuse to immobile and randomly distributed traps, which act as sinks. At short times the initial $\beta = \beta_{0i} = 1$, while at longer times, scaling of the diffusion equation [1] gives for residual relaxation in a configuration space of effective dimensionality $d*$

$$\beta_0 = d*/(d* + 2) \qquad \text{[field free]}, \tag{1}$$

giving $\beta_0 = \beta_{0s} = 3/5$ for $d* = d = 3$, often observed in field-free relaxation involving only density fluctuations. Later it was noted [13] that many values of $\beta_0$ are clustered near $\beta_{0l} = 0.43 = 3/7$, which suggested a fractal interpretation of $d*$ as

$$d* = fd \qquad \text{[fractal]}, \tag{2}$$

with $f = 1/2$ and thus $\beta = d/(d+ 4)$.

Diffusion can take place equally in two channels, only one of which is relaxational, while the other is cyclical or ineffective. The two channels could depend on short-and long-range forces, as in PG and PPG [10], with the long range (ineffective) forces being either electrostatic (OH groups), or strain (polymers, whose chain structure generates strong, long-range, H-bond mediated strain fields). [Note that relaxation of a space-filling structure by long-range forces cannot occur by diffusive exchange; it can occur only via tunneling, which is exponentially unlikely.] The two channels are nearly always weighted *equally*, a characteristic of fractal models, and the result of microscopically equal residual configurational entropies. The assumption of the latter in describing exponentially restricted glass phase space is analogous to the *a priori* equal weighting of unrestricted phase space in statistical theories of gases. It appears to be the central unifying principle that describes glassy states, whether formed experimentally by slow quenching, or in numerical simulations [2].

At this point many readers may be skeptical if they have a mental picture of glasses as extremely complex systems subject to a multitude of unspecified (and therefore uncontrolled) relaxation mechanisms. This picture is completely general, and therefore it seems to be quite appealing. In fact, however, it is fatally flawed, because it does not explain the ubiquitous occurrence of SE relaxation, with its unique distribution of relaxation times. Of course, as such pictures cannot predict SER, they are completely unable to match the amazing predictive success of the relaxation channel model with only a few discrete channels. Glasses are, in fact, able to fill space yet avoid crystallization only because they are stabilized by self-organization of internal stress fields. Self-organization defines the available relaxation channels. Of course, we do not know the details of how this occurs, but we do not need to know such details because there are only a few possibilities to be considered. Similarly, in equilibrium thermodynamics many systems that differ from each other in detail still obey the law of corresponding states. The relaxation channel model is the glassy analogue of the law of corresponding states. Much more detailed discussion of these methodological points has already been given in the lengthy review [2], which summarizes derivations of SER and many successful numerical simulations.

At this point we can return to the analysis of the birefringence data on relaxation of commercial poly(styrene-*co*-acrylonitrile) films [9]. There were two relaxation processes, with relaxation rates separated by a factor of $10^3$. The authors identified these two relaxation processes as belonging to (fast I) the main chain, and to (slow II) the phenyl side group, with $\beta_I = 0.43$ and $\beta_{II} = 0.32$. As we saw above, the value of $\beta_I = 0.43$ is exactly what we can expect from main-chain relaxation with two channels, only one of which is relaxational, while the other is cyclical or ineffective. Obviously $\beta_{II} = 0.32$ corresponds to $f_{II} = 1/3$, that is, relaxation with three channels, only one of which is relaxational, while the other two are cyclical or ineffective. What is the origin of the extra ineffective channel for the slow relaxation of the phenyl side group? It is probably the interactions with the acrylonitrile copolymer, added commercially to stabilize the film and assure microscopic homogeneity. This hypothesis suggests a series of further experiments, studying the slow relaxation as a function of added fraction of acrylonitrile. It seems likely that $f_{II} = 1/3$ only for a narrow range of compositions near the commercial value, where the macroscopic properties of the film are optimized. This theoretical prediction, if confirmed, would provide positive support for the channel model. However, it has been our experience that no amount of predictive success will satisfy all the skeptics. Like earlier unbelievers in quantum theory, one may have to wait for them to die out!

## 4. Dynamical Dielectric Relaxation

Analyses [5,6] of field-forced ionic conductive-system (doped dielectric) dispersive relaxation, measured in plane-parallel capacitor configurations, have revealed a universal frequency dependence associated with $\beta_f$. A bulk response model, called the Kohlrausch K1, involving only two free parameters with a fixed SE shape parameter, $\beta_1 = \beta_f$, fits frequency-response data excellently for a wide variety of homogeneous materials with $\beta_f$ equal to the unique value of 1/3. This model is indirectly derived from a correlation function of exact SE form involving the fixed value $\beta_1 = 1/3$. Its associated temporal response is not, however, of SE character with a constant value of $\beta_1$. Instead, separate approximate fits of the SE model to the full K1 response over limited temporal ranges show that $\beta$ approaches unity in the high frequency limit, is of the order of 0.5 for mid-range times, and reaches 1/3 only in the long-time limit.

K1 composite models with $\beta_1 = 1/3$, ones that must include dispersive electrode effects when significant, generally fit wide-frequency-range data sets appreciably better than do other models with more free parameters. The K1 part of the total response model fits the bulk part of the data exceptionally well over the entire frequency span; at high frequencies it leads to a limiting power-law log-log slope of the real part of the conductivity of $n_1 = 2/3$. In contrast to the K1 model, the K0 one is derived directly, rather than indirectly, from a correlation function of SE form with a fixed shape parameter of $\beta_0$, and its temporal response is therefore of exact SE form with a constant value of $\beta_0$ over the entire time domain. When the K0 model is used to fit the high-frequency part of K1 response with $\beta_1 = 1/3$, it leads in the high-frequency limit to the power-law exponent $n_0 = 2/3$ and to $\beta_0 = 2/3$ as well, in accordance with the well-established relation $\beta_0 + \beta_1 = 1$ [5,6].

These apparently universal shape parameter values may be explained as follows. At sufficiently low frequencies or long times, capacitor plane-parallel boundary conditions require that the current flows in one-dimensional "streams" normal to the plates, but at high frequencies the current is locally random, as expected for a fully glassy medium. Our model, as discussed below in detail, explains these two limiting fractions in terms of pinning of the current by the electric field, either globally at low frequencies, or locally at high frequencies. In addition, an entirely different theoretical model leads to the same unique values of the two limiting $\beta$ parameters, thus providing additional support for their validity and universality [6].

Results of fits of appropriate models to some experimental data sets are shown in Table I; many others are included in Refs. 5 and 6. Note that the estimated $\beta_0$ values in the table, obtained from fitting experimental data with the K0 model, are the same as the high-frequency slopes of the data derived from power-law fitting but do not reach the value of 2/3 because of the limited range of the available data. The more appropriate K1-model fits of the same data, do, however, yield high-frequency slope estimates very close to 2/3, as shown in the table. For comparison, Table I includes results for the K1 model with $\beta_1$ fixed at 1/3 during fitting as well as taken as a free fitting parameter.

Even when the experimental data involve high-frequency power-law limiting slopes near or even exceeding unity, fitting using a composite model that includes the K1 one (representing bulk response) as well as other dispersive contributions, such as that associated with partial blocking of mobile charge carriers at the electrodes, leads to bulk response best represented by the K1 with its fixed $\beta_1$ shape parameter of 1/3 and its associated high-frequency-limiting power-law slope of 2/3.

Seemingly, both ionic and electronic transport may have ranges that depend on $r^s$ with 0 < $s$ < 1 [8]. The resulting $s$ dependence of $\beta$ may be removed in a random-walk lattice model with long waiting times [16], leading to an interesting relation involving $\beta_f(d)$, namely $\beta_f(1) = \beta_f(3)/2$, first applied to frequency-domain results in [6]. This relation only appears to be satisfied in field-forced conductive dielectric dispersive relaxation data that give $\beta_{fl}(1) = 1/3$ and $\beta_{fh}(3) = 2/3$ [5,6]. The lattice model predictions agree with those of the diffusion model for $d$ =1, but disagree by ~ 10% for $d$ = 3, while it is not immediately clear what role the forcing electric field plays in distinguishing the two models at low and at high frequencies.

## 5. Fractal Dynamical Model

One can derive $\beta_{fh}(3) = 2/3$ topologically in a fractal model by postulating that in the presence of a forcing electric field, the carriers relax exponentially with respect to randomly oriented local polar coordinates. This assumption of maximally disordered local polar coordinates, compatible with space filling [17], has also been made in analyzing three-body forces in a space-filling glassy network [18]. Constraint theory has also been very successful in predicting and describing the phase diagrams of network glasses, including their glass-forming (non-crystallizing) ability, and other remarkable glass properties, such as the reversibility and non-aging compositional windows centered on average valence 12/5 [19,20]; these properties directly reflect the exponentially small phase space occupied by homogeneous glasses.

Within a local polar coordinate frame one has a radial coordinate and ($d$ − 1) angular coordinates. *A priori* in the glassy state the residual entropies associated with these coordinates should be equal, and contribute equally to exponential relaxation, represented topologically by a harmonic product

$$\exp\text{-}([t/\tau]^{1/d})^d = \exp[\text{-}t/\tau] \qquad \text{[fractal exponential]}. \qquad (3)$$

However, the $d$ configurational coordinates are not all equally effective in relaxing currents in the presence of a forcing applied electric field at a polar angle $\theta$. The field induces a uniaxially anisotropic local dynamical metric. In a general $d = 3$ geometry, motion with respect to the azimuthal coordinate $\varphi$ is irrelevant for homogeneous materials. Although there can be transient relaxation involving $\varphi$, in the steady state relaxation occurs with respect to only $r$ and $\theta$. The effective dimensionality for relaxation is thus $d^* = 2$, while the embedding dimensionality is $d = 3$, so (3) becomes

$$\exp\text{-}([t/\tau]^{1/d})^{d^*} = \exp\text{-}[t/\tau]^{d^*/d} \qquad \text{[fractal effective stretched exponential]}, \qquad (4)$$

and $\beta_{fh} = f = d^*/d = 2/3$. When currents "stream" normal to capacitor plates, transverse motion parallel to the plates is irrelevant (cyclical), giving $d^* = 1$. Now we forget about traps completely and simply assume that the residual dispersive currents are those pinned to the forcing field. In that case, the fractal or topological value of $\beta$ can only be

$$\beta_f = d^*/d = f \qquad \text{[field forced]}. \qquad (5)$$

This gives both $\beta_{fl}(1) = 1/3$ and $\beta_{fh}(3) = 2/3$, as found experimentally. This derivation has the advantage of incorporating the forcing character of the applied field explicitly in distinguishing between low and high frequencies.

One can now easily see why the hopping range power-law exponent $s$ does not appear in $\beta_f$, although it does appear in explicit algebraic models [8]; similar adjustable parameters appear in geometrical models with variable filling factors for trapping sites [7]. The algebraic models include the hopping range of the carriers, but they do not include the dynamically correlated polarization of the medium. The latter, however, arises from virtual hopping processes that have the *same* range dependence as the carrier motion; the local field corrections to the forcing field can then be screened in exactly the same way as the hopping length. But, one can ask, is the cancellation of the two factors perfect? The answer is yes, because motion in the glass is exponentially hindered, implying exponentially exact cancellation of power-law range hopping factors by power-law polarization hopping factors: after this cancellation all that can remain is the exponentially small fraction of phase space that is relaxationally effective. Therefore, we expect, as found previously for dispersive charge transport [5,6], that the 1/3 value of $\beta_{fl}$ and the 2/3 value of $\beta_0 = \beta_{fh}$ (obtained from the limiting slope of the conductivity) must be both temperature and ionic-concentration *independent*, contrary to the predictions of most geometrical models. Finally, in contrast to [16], which used a lattice model and derived only $\beta_f(1) = \beta_f(3)/2$, the present model leads to unique values for both quantities [6]. For the reader's convenience field-free and field-forced values of $\beta$ are summarized in Table II.

The distinctions made here between field-free $\beta_0$ and field-forced $\beta_f$ relaxation produce differences in $\beta$ that are sometimes numerically small (2/3 compared to 3/5), but they nevertheless imply that the two cases represent thermodynamically different phases. New low-field longitudinal magnetoconductance oscillations have been discovered in two dimensional electron glasses in a forcing microwave field [21]; these oscillations are qualitatively different from high-field de Haas-Shubnikov oscillations, and are described topologically as a new synergistic phase [22]. One can also note that dielectric hole burning in glycerol [23] is described by a dynamically heterogeneous SER model with $\beta = 0.65$, very close to our expected (field-forced) value of $\beta_f = 2/3$. Partitioning or coarse-graining is a physically appealing way to reduce phase space in non-equilibrium disordered systems [24]; however, it does not distinguish between static and dynamic inhomogeneities, fractal dimensionalities, or forcing effects.

## 6. Anisotropy of Cuprate Photoinduced (Super)conductivity

While there are many examples of intrinsic topological relaxation, the most demanding application is probably to photoinduced (super)conductivity in the cuprates (specifically $YBa_2Cu_3O_x$ at the metal-insulator transition, x = 6.4) [2,25]. Here the laser field clearly is forcing the development of a field-pinned internal filamentary structure upon a precursive percolative structure, so both field-free and field-forced mechanisms are contributors to this remarkable effect: which makes the greater contribution? The superconductive cuprates are layered pseudoperovskites, with in-plane conductivities 10-100 times larger than perpendicular conductivities. The phenomena exhibit SER, so we can use the anisotropy of observed values of $\beta$, shown in Table III, to answer this question. The in-plane low-temperature limit agrees well with $\beta_0 = 3/5$, suggesting that the internal structure develops mainly through atomic density relaxation. The systematic difference $\delta\beta \sim 0.05$ between in-plane and out-of-plane $\beta$ values suggests that the in-plane rearrangement is more nearly completed (with applied fields shorted out by the large planar conductivity), while the plane-normal relaxation still retains some memory of the insulating state, and resembles the high-frequency limit $\beta_{fh} = 2/3$ of field-forced relaxation. Thus this small difference in $\beta$ appears to be significant quantitatively, and it is in accord with our intuitive expectations regarding the photoinduced formation of weak, low conductivity interplanar links (traps in the insulating layers).

## 7. Conclusions

The dynamic and thermodynamic properties of glasses are rich and diverse; they appear in at least twelve different categories, and all seem to be related through $\beta$ [26, 27]. In our experience in fitting one category, dispersive transport [5,6], we have found that $\beta_f$ (which describes the shape of the relaxation-time distribution) is much more stable than is the SE characteristic relaxation time $\tau$ itself (which is discussed more often, especially in the context of percolative models), and can be used to test power-law assumptions regarding frequency or time ranges outside those studied experimentally. For statically homogeneous samples one can have much more confidence in data fits that yield magic fractions for $\beta$, and therefore more confidence in chemical trends of other less stable parameters (such as $\tau$) that depend implicitly on $\beta$ as well as other "hidden" parameters. In emerging fields of materials science (such as cuprates), the materials that yield magic fractions for $\beta_0$ are consistently of higher quality [2]. By explaining the topological origins of β without the use of adjustable parameters, the present theory may provide useful guides to more consistent theories of all the properties of glasses.

*Postscript.*  A complex geometrical (hyperspherical) fractal model of random walks on percolation clusters [28] yields SER exactly, showing that SER is the result of phase-space compression.  However, the values of β range from ~ 0.7 ($d = 2$) to 1/3 ($d \rightarrow \infty$), in disagreement with our results and experiment.  Note that the percolation clusters, although exponentially unlikely, are still *static* geometrical constructs, and that these constructs need not be internally consistent with glassy relaxation *dynamics*.

After [2] was published, skeptics who picture glasses as extremely complex systems subject to a multitude of unspecified (and therefore uncontrolled) relaxation mechanisms suggested privately that the relaxation channel model provides only an incomplete description of glassy relaxation at the microscopic level.  The theory had been found to fit relaxation of the intermediate scattering function S($\mathbf{Q}$,t) only at $\mathbf{Q} = \mathbf{Q}_1$, where $\mathbf{Q}_1$ is the position of the largest peak. A more complete description, they said, would include a function β($\mathbf{Q}$), which would oscillate, much as S($\mathbf{Q}$,0) does.  This conjecture has recently been tested [29] with spin polarized neutron scattering studies of S($\mathbf{Q}$,t) in polybutadiene (PB) and polyisobutylene (PIB) in the temperature range above the glass transition dominated by segmental relaxation.  Even in this range in both cases the result was that β is fixed and independent of $\mathbf{Q}$.  These results provide further validation of the relaxation channel model and greatly enhance the significance of data obtained by spin polarized neutron scattering as measures of intrinsic relaxation dynamics that are independent of the probe neutron momentum.

Table I. Complex nonlinear least squares frequency-response fitting results using K1 (proper high frequency limit) and K0 (uncorrected high frequency limit) models [4,5]. We deal here with $\beta_f$ quantities, but designate them by $\beta_k$, with the subscript $k$ specifying the model used for fitting. The $\beta_1 = 1/3$ values listed below were held constant during fitting. Here $100S_F$ is the percentage relative standard deviation of a fit. The high frequency limiting slope of the real part of the conductivity is $(1 - \beta_1)$ for a K1 model and is the same in the limit when the K0 model is used to fit very-wide-range K1 synthetic data; so then $\beta_0 = 1 - \beta_1$.

| Material | $T$ (K) | Fit models | $100S_F$ | $\beta_k$ | High frequency limiting slope |
|---|---|---|---|---|---|
| | | | | | |
| $0.88ZrO_2 \bullet 0.12Y_2O_3$ (single crystal) | 503 | K1 | 0.49 | 1/3 | 2/3 |
| | | K1 | 0.43 | 0.32 | 0.68 |
| | | K0 | 0.76 | 0.49 | 0.49 |
| $0.03Na_2O \bullet 0.97GeO_2$ | 571 | K1 | 1.4 | 1/3 | 2/3 |
| | | K1 | 1.3 | 0.36 | 0.64 |
| | | K0 | 2.3 | 0.55 | 0.55 |
| $Na_2O \bullet 3SiO_2$ | 303 | K1 | 0.55 | 1/3 | 2/3 |
| | | K1 | 0.56 | 0.33 | 0.67 |
| | | K0 | 0.67 | 0.56 | 0.56 |
| $Li_2O \bullet Al_2O_3 \bullet 2SiO_2$ | 297 | K1 | 2.2 | 1/3 | 2/3 |
| | | K1 | 2.2 | 0.34 | 0.66 |
| | | K0 | 3.0 | 0.56 | 0.56 |

Table II.  Values for β predicted for molecular and ionic relaxation depend on the range of forces (short or both long and short) for the former, and on whether or not one includes the correct high-frequency limiting slope behavior (entered here as *) in transforming the frequency data of the latter to the temporal regime. When the latter is included, $\beta_{fl}$ is given as 1/3, but when it is omitted, (values) close to 1/2 are found (Table I) for usual limited-range data.

| Excitation | Notation | Internal | External | $d$ | $f$ | β | Ref. | Year |
|---|---|---|---|---|---|---|---|---|
| Molecular | $\beta_0$ | short | none | 3 | 1 | 3/5 | 12 | 1973 |
| Molecular | $\beta_0$ | long | none | 3 | 1/2 | 3/7 | 13 | 1994 |
| Ionic | $\beta_{fl}$ | none | low | 3 | 1/3 | 1/3 (~1/2) | 6 | 2004 |
| Ionic | $\beta_{fh}$ | none | high | 3 | 2/3* | 2/3 | 6 | 2004 |

Table III.  SER parameters for photoexcitation of resistivities in layered YBCO [19]. The theory, discussed in the text, refers to an ideal sample without ferroelastic fluctuations, which may reduce both β's by 5-10%.

| T(K) | β (in plane) | β (out of plane) |
|---|---|---|
| 275 | 0.54 | 0.57 |
| 200 | 0.55 | 0.59 |
| 125 | 0.56 | 0.61 |
| 50 | 0.58 | (noisy) |
| Theory | 0.60 | 0.67 |